\documentstyle[preprint,aps]{revtex}
\begin{document}
%
%
\title{
On the Observation of Top Spin Correlation Effect at Tevatron
}
\author{
Darwin Chang$^{(1,2)}$,
Shih-Chang Lee$^{(2)}$ and
Alexei Soumarokov$^{(2)}$ }
\address{
$^{(1)}$Physics Department,
National Tsing-Hua University, Hsinchu 30043, Taiwan, R.O.C.}
\address{
$^{(2)}$Institute of Physics, Academia Sinica, Taipei, R.O.C.}
\maketitle

\begin{abstract}
We propose to observe the top spin correlation effect at Tevatron through
the measurement of the correlated asymmetries of the charged lepton momenta
using the dilepton decay events of top and anti-top quark pairs.
The possibility of reconstructing the events and observing the
asymmetries at 3$\sigma$ level with projected luminosity of the Tevatron
for Run II is demonstrated by simulation.
The effect can provide the first direct measurement of the spin of a quark.
\end{abstract}


Both CDF and D0 have recently announced the observation of top quark
\cite{cdf} .  The published central value of the top quark mass is well
above twice the $W$ boson mass.  The heaviness of top quark offers an
unique
chance to observe the $t\bar{t}$ spin correlation effects through their
decay products because the top quark is expected to decay before it has
time to hadronize\cite{bigi,clt}.  Despite many studies in the
literature\cite{clt,barger}, the
possibility of observing the top spin correlation at Tevatron with the
projected luminosity for Run II is yet to be demonstrated.

In this letter, we propose to measure the charged lepton correlated
asymmetries in the top dilepton events.  We shall demonstrate that it is
possible to observe such asymmetries at $3\sigma$ level with Tevatron II.
The lepton correlated asymmetry, $A_P$, with
respect to a plane $P$ passing through the interaction point is defined
as follows.  Let $\vec{l}_1$
be the momentum of the charged lepton associated with top decay
evaluated in the top rest frame.
Similarly, let $\vec{l}_2$ be that associated with anti-top decay
evaluated also in its own rest frame.
Then, define $A_P = (N_+ - N_-)/(N_+ + N_-)$ where
$N_+$ is the number of
events in which both
$\vec{l}_1$ and $\vec{l}_2$
lies on the same side of $P$ while
$N_-$ is the number of
events with both
$\vec{l}_1$ and $\vec{l}_2$
lying on the opposite side of $P$.
By choosing different $P$, one can construct different asymmetries.
These asymmetries vanish when the effects of both top spin and W spin are
ignored (or averaged over) and
they remain small even when the W spin effects are included.  This
property makes them the ideal candidates for the observation of the
$t\bar{t}$ spin correlation effects.

In order to measure the lepton correlated asymmetries it is essential to
make full reconstruction of the top dilepton events.  We assume that
the top mass will be well measured in Tevatron Run I and Run II.  For
dilepton candidate events, we further assume that the missing transverse
momentum measured is equal to the sum of the missing transverse momenta of
the two neutrinos associated with the dilepton.  Contribution to the
missing transverse momentum from other neutrinos in the event reduces the
efficiency of the reconstruction and lowers the signal-to-noise ratio but
does not spoil the observability of the asymmetries we consider here.

The analysis of the productions and the decays of $t\bar{t}$ in
hadronic collider with and without spin correlation effect can be found in
the literature\cite{clt,barger} with various degrees of sophiscations.
We shall use the simple analytic differential cross sections for
$q \bar{q}$ and $g g \rightarrow
t \bar{t} \rightarrow b W^+ \bar{b} W^- \rightarrow b \bar{l} \bar{\nu_l}
\bar{b} l' \nu_{l'}$
provided in Ref.\cite{clt}.
Using these formulas, we simulated the $t\bar{t}$ production and decay
employing the event generator PYTHIA 5.7.

The algorithm of our reconstruction of the top dilepton events goes as
follows.  We choose two arbitrary directions in the transverse plane as the
candidate directions of the transverse momenta for the two neutrinos.
The momenta of the two neutrinos are then fixed by the four on-shell
conditions of the top quarks and the $W$ bosons.  In general, four
candidate solutions are obtained this way.  One then compute the absolute
value of the difference,
$\Delta E_{\perp}$,
between the sum of the
transverse momenta of the two neutrinos and the measured missing
transverse momentum of the event.  Varying the two candidate directions
in the transverse plane, we can choose as our best solution the one with
minimum $|\Delta E_{\perp}|$.
Once the neutrinos are reconstructed, top momenta can be calculated.
Invariably, our reconstruction does not always produce fully the true
kinematics of the collision, however, for our purpose, it is sufficient
that the reconstruction produces the true top direction with enough
efficiency.
We checked that the reconstructed top direction highly peaked around the
true top direction.

In principle, one can also look for the $W$ correlated asymmetries for
the two $W$'s from $t\bar{t}$ decays in the reconstructed top dilepton,
lepton-jet and jet-jet events.  The predicted $W$ asymmetries can be
easily obtained from Ref.\cite{clt}.  They are no more than a few per
cent and about an order of magnitude smaller than some of the lepton
correlated asymmetries.  The enhancement of the lepton correlated
asymmetries over the $W$ correlated asymmetries\cite{cls} is one feature
that favors the observation of top spin correlation effects through the
dilepton channel.

It is interesting to note that the predicted correlated asymmetries for
the neutrinos vanish\cite{cls}.  This lack of symmetry between the
properties of charged
lepton and neutrino reflects exactly the $V-A$ coupling of the weak
charged current interaction.  Therefore, the observation of a nontrivial
lepton correlated asymmetry and vanishing neutrino correlated asymmetries
can by itself constitute a check on the $V-A$ coupling of the charged
weak current between $t$ and $b$.
Unfortunately, as we shall demonstrate, the typical trigger for missing
$E_{\perp}$
used to select $t\bar{t}$ candidate events introduces bias so
that the neutrino correlated asymmetries are no longer vainshing when the
trigger is turned on.  However, one may still hope that by understanding
this trigger-induced effect better one can derive some direct experimental
constraint on $g_A/g_V$ coupling of top quark.

The lack of symmetry between charged lepton and neutrino in their
asymmetries
also implies that in order to look for lepton-quark correlated asymmetries
in the top lepton-jet events, one should distinguish between the up-type
and the down-type jets in the W decay.  This is difficult, if not
impossible.
Since the lepton-jet events are much more abundant than the dilepton
events, the possibility of observing the lepton-quark asymmetries may
still deserve further study.

In hadron colliders, $t\bar{t}$ are produced either by quark-antiquark
($q\bar{q}$) annihilation or by gluon-gluon ($gg$) fusion.  The lepton
correlated asymmetries for these two cases typically have opposite sign
and with different magnitudes\cite{cls}.
At Tevatron energy,
the number of $q\bar{q}$ produced top events are roughly eight times that
of the $gg$ produced events.
This ratio decreases with energy and, at LHC energy,
the number of $gg$ produced top events are roughly four times that
of the $q\bar{q}$ produced events.
For the asymmetries we considered, we found that the substantial
asymmetries that can be observed at Tevatron become much smaller at the
expected LHC collider energy.
This is the result of accidental cancellation between the
contribution of the $q\bar{q}$ production channel and that of the $gg$
production channel.  We have checked that the contribution due to the
$q\bar{q}$ channel alone is indeed not so small.
This cancellation could be a generic feature which may
make LHC unfavorable machine to look for top spin correlation effect.
For the same reason, the next linear colliders should be an ideal machine
for observing such correlation due to the absence of such cancellation.

In this note we shall only discuss the asymmetries with respect to the
following planes defined in the $t\bar{t}$ center of mass frame: (1).
$t\bar{t}$ production plane (defines asymmetry $A_1$); (2). the plane
perpendicular to the production plane and contains the top (asymmetry
$A_2$); (3). the plane perpendicular to the two previous planes (asymmetry
$A_3$); (4). the plane normal to the beam direction (asymmetry $A_4$).
These planes are chosen as samples to demonstrate the possibilies only.
The optimal choice will depend on the details of the detectors and
clearly need further study\cite{cls}.

The typical trigger for top dilepton events\cite{cdf}, namely,
$p_T \geq 20 GeV$ for leptons,
$p_T \geq 10 GeV$ for $b$-jet, missing transverse energy
$\not{\!\!E}_{\perp} \geq 25 GeV$ and rapidity $|\eta| \leq 2$, was applied.
Afterwards, the reconstruction algorithm we described earlier was carried
out.
A typical hadron calorimeter energy resolution of $70\%/\sqrt{E}$ was used
as b - jet energy smearing and the missing transverse energy was Gaussian
smeared with a $15\%$ standard deviation.
The effect of including contribution of other neutrinos
in an event to the missing transverse energy
was investigated\cite{cls}.  To isolate the effect of the bias originated
from
reconstruction algorithm to the asymmetries, we also studied the case
with the event reconstruction turned on but with the trigger turned off.
Similarly we also studied the case
with the event reconstruction turned off but with the trigger turned on.
The effect of energy smearing is also investigated.
In each case, we computed the lepton correlated asymmetries,
the neutrino correlated asymmetries and the $W$
correlated asymmetries with respect to the planes described earlier.
In Table \ref{t1}, we present the measured correlated asymmetries
$A_1, A_2, A_3$ and $A_4$
for the four cases:
(I). trigger off, reconstruction off;
(II). trigger on, reconstruction off;
(III). trigger off, reconstruction on;
(IV). trigger on, reconstruction on;
(V). trigger on, reconstruction on and with energy smearing included.
The corresponding asymmetries
when the top spins were ``uncorrelated''
(that is, their spins are summed over in their productions and
averaged over in their decays,)
are given in brackets.
The average
values and the standard deviations of the asymmetries were extracted
directly from the simulated data.  When both top and $W$ spins were
``uncorrelated'',
as in most standard event generator packages, we verified that all the
asymmetries vanish if the trigger and event reconstruction were turned off.
As one can see in the table, the asymmetries measured by charged
leptons are generally larger than the asymmetries measured by neutrinos
and $W$ bosons.
The systematic effects of trigger and reconstruction are quite obvious in
the case of neutrino asymmetries as one would expect from the large
missing $E_{\perp}$ cut as well as the contributions from other neutrinos
in the event.  Due to space, a detailed discussions of the
various effects and their origins will be given elsewhere.

{}From the results of these simulation, we conclude that the lepton
correlated asymmetries arising from the $t\bar{t}$ spin correlation
can be observed at 3$\sigma$ level with a few hundred top dilepton
events.
This is certainly reachable with the projected luminosity of the
Tevatron for Run II, with improved detector resolution and acceptance
expected for both CDF and D0 detectors upgrades and perhaps with improved
algorithms for identifying the top dilepton events.

There are also plenty of rooms for improvement on our analysis of the
correlated asymmetries.  Even though trigger and reconstruction show
little systematic effects for the lepton asymmetry, they do shift the
neutrino and the $W$ asymmetries by non-negligible ammounts. A
quantitative understanding of these effects may allow us to tune the
trigger selection and reconstruction algorithm for the observation of
the asymmetries.  It is well known that discrete ambiguities exist in
general in the reconstruction of top dilepton events.  A detailed
quantitative study of its effect on the asymmetries could precipitate an
improvement on reconstruction algorithm for better efficiency and better
signal-to-noise ratio.
We have been quite casual in choosing the planes to define the
asymmteries and in choosing the combinations of these asymmetries to
measure.
One may wonder if
there is an optimal choice of plane or combinations of planes that can
maximize the observability of the correlation effect.  One may also
wonder if there are choices that will enhance or suppress the
contribution of $q\bar{q}$ annihilation channel relative to the $gg$
fusion channel.

Before conclusion, we wish to point out still another
way of measuring the asymmetry.
The asymmetries are functions of the orientations of the defining plane,
${\rm P}$, passing through
the collision point.  Hence each asymmetry (charged lepton's, neutrino's
or $W$'s) can be considered
as a function on a sphere and analyzed by multipole expansion.  As an
illustration, in Fig. 1, we plot the charged lepton asymmetry, without
trigger and reconstruction effects, as a function of the plane ${\rm P}$
for both the case with top spin correlation (Fig. 1a) and the case without
(Fig. 1b).  With improved reconstruction algorithm and detector
resolution, this can potentially be a very effective way of searching for
the top spin correlation effects\cite{cls}.

In conclusion, top spin correlation is certainly one of the most
interesting top physics
to be uncovered.  It can provide a direct observation of the spin $1/2$
character of top quark (which we have not been able to do for the lighter
quarks) and can potentially test the $V-A$ character of the weak charged
current associated with top.  We have clearly demonstrated the possibility
of observing this effect at Tevatron Run II.  The fact that this effect
may be even harder for LHC to measure should make the task more important
for Tevatron.  A detailed account of our study will be presented
elsewhere\cite{cls}.

\acknowledgments

We would like to thank Paul Turcotte, Vernon Barger, Jim Ohnemus, Stephen
Parke, Chris Quigg for useful discussions. This work is supported
by the National Science Council of Republic of China
under grants
NSC85-2811-M008-001(for S.L and A.S.),
NSC85-2112-M007-029, NSC85-2112-M007-032(for D.C.).

%
%

\onecolumn

\begin{table}
\caption{
The measured correlated asymmetries $A_i(l)$'s, $A_i(\nu)$'s
and $A_i(W)$'s
for the four cases:
(I). trigger off, reconstruction off;
(II). trigger on, reconstruction off;
(III). trigger off, reconstruction on;
(IV). trigger on, reconstruction on;
(IV). trigger on, reconstruction on with energy smearing.
The corresponding asymmetries when the top spins were ``uncorrelated''
are given in brackets.  All the values are in unit of percentage.
The total number of simulated events is 100000.
}
\label{t1}
\begin{tabular}{cccccc}
$$       &         $I$          &    $II$         & $III$        & $IV$       &
$V$
\\
\hline
$A_1(l)$ & $  3.99(-.11)$&$  5.90( .58)$&$  4.16(  .68)$&$  6.42( 1.94)$&$
5.50(1.14)$
\\
$A_2(l)$ & $-11.78( .41)$&$-11.18(-.55)$&$-10.30( -.30)$&$ -7.98(  .12)$&$
-7.58(-.06)$
\\
$A_3(l)$ & $ -9.65(-.09)$&$-8.68(-1.34)$&$ -8.00(-1.22)$&$ -6.80( -1.4)$&$
-6.64(-1.26)$
\\
$A_4(l)$ & $-20.34(
.18)$&$-17.64(-.13)$&$-16.68(-1.86)$&$-13.72(-1.66)$&$-13.30(-2.00)$
\\
$A_1(\nu)$&$  0.54(-.68)$&$  9.50(8.86)$&$  5.44( 4.70)$&$ 12.24(11.80)$&$
9.02( 8.48)$
\\
$A_2(\nu)$&$ -1.95(-.06)$&$   .60(1.68)$&$  1.80( 2.20)$&$  2.16( 2.42)$&$
2.56( 2.52)$
\\
$A_3(\nu)$&$ -1.45(-.09)$&$  3.62(4.30)$&$ -1.94(-1.54)$&$  2.64( 2.82)$&$
.74( 1.22)$
\\
$A_4(\nu)$&$ -3.41(-.37)$&$ -1.84(-.13)$&$   .33(  .27)$&$  -.02(  .24)$&$
.84( 1.02)$
\\
$A_1(W)$ & $   .79( .16)$&$  3.24(2.32)$&$  1.92( 1.34)$&$  3.94( 3.18)$&$
3.70( 2.92)$
\\
$A_2(W)$ & $ -1.98(-.51)$&$  -.48( .33)$&$ -3.34( -.12)$&$ -3.14(  .58)$&$
-2.96(  .58)$
\\
$A_3(W)$ & $ -2.70(-1.46)$&$  .56( .42)$&$ -3.70(-1.44)$&$ -3.64(-1.32)$&$
-4.60(-2.34)$
\\
$A_4(W)$ & $ -3.66( -.91)$&$-3.32(-.04)$&$ -7.12(-1.68)$&$ -6.58( -.72)$&$
-7.57(-1.62)$
\\
\end{tabular}
\end{table}

\figure{Fig. 1a.
Charged lepton asymmetry $A_P(l)$ as a function of the orietation of
the defining plane ${\rm P}$ for the case with top spin correlation.
Here, $\theta = 0$ represents the plane perpendicular to the beam
direction while $\theta = \pi/2,\,\, \phi = 0$
represents the production plane.}

\figure{Fig. 1b.
Same as Fig. 1a but for the case when top spin are ``uncorrelated''.}


\begin{references}
%
\bibitem{cdf}  F. Abe {\it et al.} Phys. Rev. Lett. {\bf 73}, 225, 1994;
F. Abe {\it et al.} Phys. Rev. D{\bf 50}, 2966, 1994.
\bibitem{bigi}
I. Bigi and H. Krasemann, Z. Phys. {\bf C7} 127 (1981);
I. Bigi, Y. Dokshitzer, V. Khoze, J. Kuhn and P. Zerwas,
Phys. Lett. B {\bf 181} 157 (1986).
\bibitem{clt}
D. Chang, S.-C. Lee and P. Turcotte,
Preprint hep-ph-9508357(1995).
\bibitem{barger}
R. Kleiss and W.J. Stirling,
Z. Phys. C {\bf 40} 419 (1988);
V. Barger, J. Ohnemus and R.J.N. Phillips,
Int. J. Mod. Phys. A {\bf 4} 617 (1989);
G.L. Kane, G.A. Ladinsky, C.-P. Yuan,
Phys. Rev. D {\bf 45}, 124, (1992).
\bibitem{cls}
D. Chang, S.-C. Lee and A. Soumarokov,
in preparation.
\end{references}
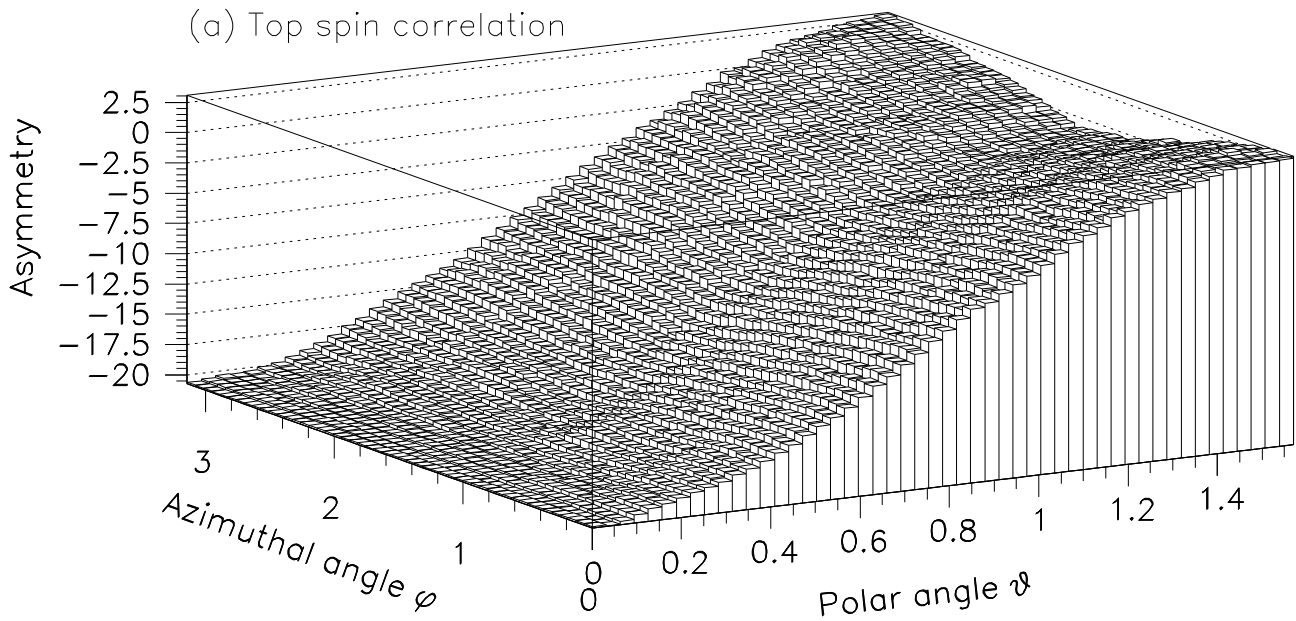
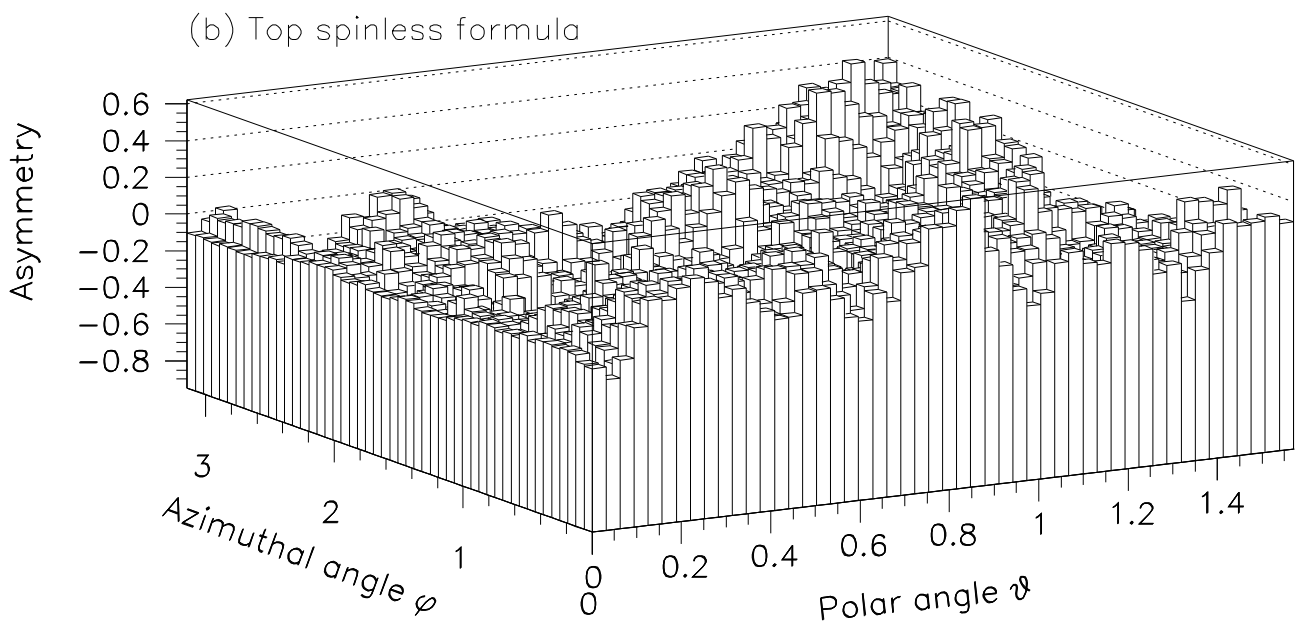
\end{document}